\documentclass[%
aps,%
showpacs,%
twocolumn,%
floatfix,%
superscriptaddress,%
nofootinbib,%
preprintnumbers,%
showkeys
]{revtex4}

\usepackage[dvips]{graphicx}
\usepackage{color}
\usepackage{epsfig}
\usepackage{amsmath}
\usepackage{amssymb}
\usepackage{booktabs}
\usepackage{array}
\usepackage{afterpage}

\def\stau{\tilde{\tau}}
\def\neutralino{\tilde{\chi}^0}

\preprint{STUPP-07-190}



\begin{document}

\title{Possible solution to the $^7$Li problem by the long lived stau}

\author{Toshifumi Jittoh}
\email{jittoh@krishna.th.phy.saitama-u.ac.jp}
\affiliation{Department of Physics, Saitama University, 
        Shimo-Okubo, Sakura-ku, Saitama, 338-8570, Japan}
\author{Kazunori Kohri}
\email{k.kohri@lancaster.ac.uk}
\affiliation{Physics Department, Lancaster University LA1 4YB, UK}
\author{Masafumi Koike}
\email{koike@krishna.th.phy.saitama-u.ac.jp}
\affiliation{Department of Physics, Saitama University, 
        Shimo-Okubo, Sakura-ku, Saitama, 338-8570, Japan}
\author{Joe Sato}
\email{joe@phy.saitama-u.ac.jp}
\affiliation{Department of Physics, Saitama University, 
        Shimo-Okubo, Sakura-ku, Saitama, 338-8570, Japan}
\author{Takashi Shimomura}
\email{takashi@krishna.th.phy.saitama-u.ac.jp}
\affiliation{Department of Physics, Saitama University, 
        Shimo-Okubo, Sakura-ku, Saitama, 338-8570, Japan}
\author{Masato Yamanaka}
\email{masa@krishna.th.phy.saitama-u.ac.jp}
\affiliation{Department of Physics, Saitama University, 
        Shimo-Okubo, Sakura-ku, Saitama, 338-8570, Japan}

\begin{abstract}
Modification of standard big-bang nucleosynthesis is
considered in the minimal supersymmetric standard model to
resolve the excessive theoretical prediction of the abundance
of primordial lithium 7.
We focus on the stau as a next-lightest superparticle, which is long lived 
due to its small mass difference with the lightest superparticle. 
It provides a number of additional decay processes of $\mathrm{^{7}Li}$ and $\mathrm{^{7}Be}$. 
A particularly important process is the internal conversion in the stau-nucleus bound state, 
which destroys the  $\mathrm{^{7}Li}$ and $\mathrm{^{7}Be}$ effectively. 
We show that the modification can lead to a prediction consistent with the observed abundance of
$\mathrm{^{7}Li}$. 
\end{abstract}

\maketitle

\section{Introduction}

The theory of big-bang nucleosynthesis (BBN) has been successful in
predicting the abundance of light elements in the Universe from a
single parameter, baryon-to-photon ratio $\eta$.
The recent results of the Wilkinson microwave anisotropy probe (WMAP)
experiment~\cite{Spergel:2006hy}, however, put this theory into
challenge.
The extraordinarily precise results from WMAP are put together with 
the standard BBN (SBBN) to predict the abundance of $\mathrm{^{7}Li}$ to
be $(4.15^{+0.49}_{-0.45}) \times 10^{-10}$\cite{Coc:2003ce} if we
adopt $\eta = 6.1 \times 10^{-10}$ (68 $\%$
C.L.)~\cite{Spergel:2006hy}.
This prediction is inconsistent with the observation of metal-poor
halo stars which implies $(1.23^{+0.32}_{-0.25}) \times 10^{-10}$
\cite{Ryan:1999vr} reported by Ryan \textit{et al}.
\cite{Cyburt:2003fe}.
The inconsistency persists even if we adopt the recent
observations, which give the less restrictive constraint of
$(2.19^{+0.30}_{-0.26}) \times 10^{-10}$ \cite{Bonifacio:2002yx} and
$(2.34^{+0.35}_{-0.30}) \times 10^{-10}$ \cite{Melendez:2004ni}.
This discrepancy can be hardly attributed to the correction of the
cross section of nuclear reaction \cite{Cyburt:2003ae,Angulo:2005mi},
and astrophysical solutions are pursued \cite{Korn:2006tv}.

Another interesting approach to this problem would be to consider
the effects induced by new physics beyond the standard model (SM).
Exotic particles which interact with nuclei will open new channels to
produce and destroy the nuclei, giving a potential solution to the
$\mathrm{^{7}Li}$ problem.
In this paper we investigate a possibility that the interaction is
initiated by a formation of the bound state of exotic negatively charged massive
particles (CHAMPs) and a nucleus.
(For the other solutions, see
\cite{ichikawa-Li7,Jedamzik-Li,Kohri:2005wn}.)

So far the bound-state effects by CHAMPs have been attracting many
interests
\cite{cahn:1981,bib:CBBNold,Pospelov:2006sc,Kohri:2006cn,bib:CBBN,yanagida_catalysed}.
For doubly charged particles, see also
Refs~\cite{DoubleCharge}.  
In particular, a significant enhancement of a $^{6}$Li-production rate
through $\mathrm{^{4}He} + \mathrm{D} \to \mathrm{^{6}Li} + \gamma$ by
the bound state with $\mathrm{^{4}He}$ was reported
\cite{Pospelov:2006sc} for the first time and recently confirmed
\cite{Hamaguchi:2007mp}.
This hinders the compatibility between
particle physics models and BBN \cite{Kawasaki:2007xb}.

In addition, some nonstandard effects on the abundance of
$\mathrm{^{7}Li}$ and $\mathrm{^{7}Be}$ were also considered in
Ref.~\cite{Kohri:2006cn} and more recently in Ref.~\cite{Bird:2007ge}.
Introducing the CHAMPs with the mass of electroweak scale, the authors
in Ref.~\cite{Bird:2007ge} newly considered several destruction
channels of $\mathrm{^{7}Be}$ nuclei through the trapping of the CHAMPs
to show that the abundance of the CHAMPs needs to be larger than 0.02
per baryon and that their lifetime has an allowed window between $1000$
 and $2000 \, \mathrm{sec}$.

We put the CHAMP BBN scenario in the minimal supersymmetric standard
model (MSSM) with the conservation of $R$ parity.
MSSM doubles the particle content of the SM by introducing 
superparticles, which can accommodate the CHAMPs.
The CHAMPs need a lifetime long enough to sustain the sufficient
abundance at the time of nucleosynthesis.
Although the $R$ parity conservation stabilizes the lightest
superparticles (LSPs), the observational constraints exclude charged
superparticles as a candidate for LSPs, which is usually considered to
be neutralinos $\tilde{\chi}^{0}$ or gravitinos.
A possible candidate of CHAMPs is the next-lightest superparticle
(NLSP) with electric charge, which can have a long lifetime by assuming
a small mass difference from the LSP \cite{Jittoh:2005pq}.

We assume in the present paper that the LSP is a neutralino and the
NLSP is a stau $\tilde{\tau}$, the superpartner of tau
lepton $\tau$.
The staus can decay into neutralino LSP with the hadronic current,
through which they also interact with the nuclei.
The gravitino LSP, on the other hand, does not couple with hadronic
current.
We consider the bound state of $\mathrm{^{7}Be}$ and
$\tilde{\tau}^{-}$ in the early Universe and the subsequent decay
chain of nucleus $\mathrm{^{7}Be} \to \mathrm{^{7}Li} \to
\mathrm{^{7}He}$ due to the interactions of the two.
The $\mathrm{^{7}He}$ nuclei rapidly decay into $\mathrm{^{6}He}$
nuclei, which are effectively stable in the considered time scale.
With the freedom of the mass of stau $m_{\tilde{\tau}}$ and its
lifetime $\tau_{\tilde{\tau}}$, we search for the possible solution to
the $\mathrm{^{7}Li}$ problem that are phenomenologically acceptable.

This paper is organized as follows.  In Sec. \ref{sec.channels} we
overview some new decay channels by stau and estimate their
lifetimes.  Here we will see that stau-nucleus bound states play an
important role in $^7$Be/$^7$Li reducing processes.  In
Sec. \ref{sec.numercal_calculation} we numerically calculate
primordial abundances of light elements while taking into account the
new channels.  Then we will see the possible solution of the $^7$Li problem.
Finally, summarization is made in Sec. \ref{sec.summary}.

\section{The Destruction of $^7$B\lowercase{e}/$^7$L\lowercase{i} in
  MSSM}
\label{sec.channels}

\subsection{Elementary interactions of the staus}

We consider a modification of the SBBN scenario under the MSSM.
MSSM introduces a set of superparticles as the partners of the
particle appearing in the standard model (SM).
The superparticles interact with the standard particles and thus
introduce additional decay channels of $\mathrm{^{7}Be}$ and
$\mathrm{^{7}Li}$ to the standard BBN theory.
The additional channels give a possible solution to the problem
where the theoretical prediction of the abundance of $\mathrm{^{7}Be}$
and $\mathrm{^{7}Li}$, or collectively $\mathrm{^{7}Be/^{7}Li}$,
exceeds the observational results by a factor of $\sim 2 \textrm{--}
3$.
We consider in the present paper that the destruction of primordial
$\mathrm{^{7}Be/^{7}Li}$ nuclei is due to their interaction with
the negatively charged staus $\tilde{\tau}^{-}$, the superpartner of the tau
lepton $\tau^{-}$, which we identify as the next-lightest
superparticle (NLSP).
The mass eigenstate of the stau is given by the linear combination of the
left-handed stau $\tilde{\tau}_{\textrm{L}}$ and the right-handed
staus $\tilde{\tau}_{\textrm{R}}$ as
\begin{equation}
  \tilde{\tau}
  = \cos \theta_{\tau} \tilde{\tau}_{\textrm{L}}
  + \sin \theta_{\tau} \mathrm{e}^{-\mathrm{i} \gamma_{\tau}}
    \tilde{\tau}_{\textrm{R}},
\end{equation}
where $\theta_{\tau}$ is the left-right mixing angle and
$\gamma_{\tau}$ is the CP-violating phase.

Staus have attractive features when considering the destruction of
$\mathrm{^{7}Be/^{7}Li}$.
First, staus have a negative charge and can form a bound state with
nuclei so that they interact efficiently.
Second, staus couples with the hadronic current $J^{\mu}$, through
which they interact with nuclei as we see below.
Third, staus can be abundant at the time of BBN.  They can acquire 
the sufficientlty long lifetime when the staus and LSPs, which we assume 
as the nutralinos, have a mass difference tiny enough.

\begin{figure}
  \begin{center}
    \includegraphics[width=80mm]{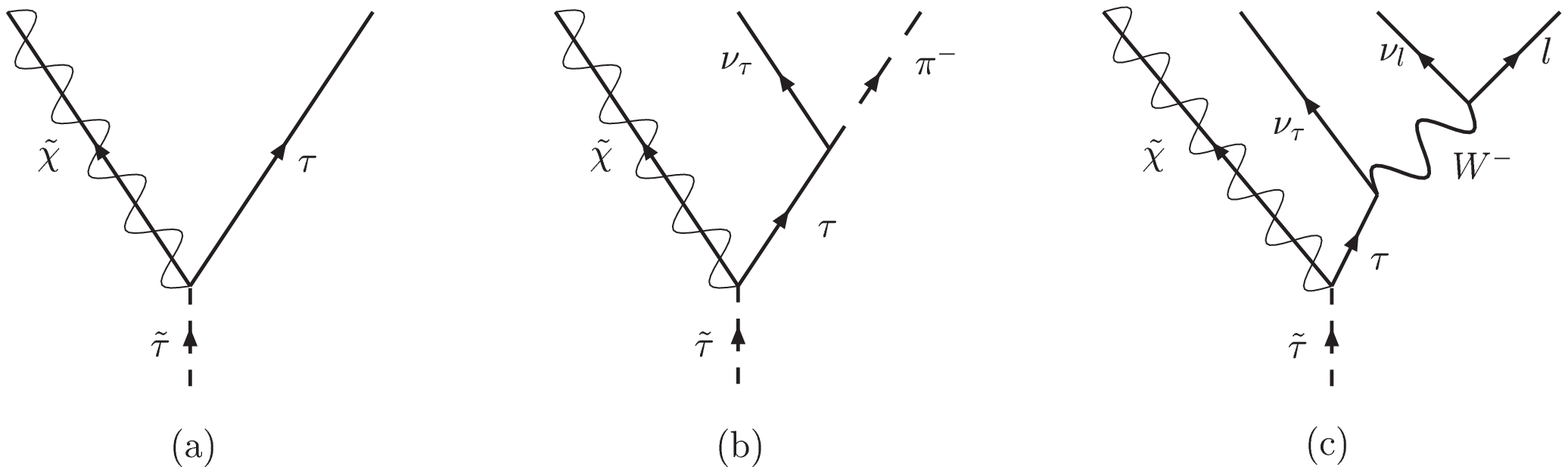}
    \caption{Feynmann diagrams of the decay of staus. 
                        (a) $\stau \to \tau \neutralino$, 
                        (b) $\stau \to \pi \nu_\tau \neutralino$,
                        and (c) $\stau \to l \nu_l \nu_{\tau} \neutralino$.}
    \label{free_diagrams}
  \end{center}
\end{figure}

The interaction of staus is described by the Lagrangian
\begin{equation}
\begin{split}
  \mathcal{L}_{\textrm{int}}
  & =
    \tilde{\tau}^{\ast} \overline{\tilde{\chi}^{0}}
    (g_{\textrm{L}} P_{\textrm{L}} + g_{\textrm{R}} P_{\textrm{R}}) \tau 
  + \frac{4 G}{\sqrt{2}} \nu_{\tau} \gamma^{\mu} P_\textrm{L} \tau J_{\mu}
  \\ & \quad
  + \frac{4 G}{\sqrt{2}}
    (\bar{l} \gamma ^{\mu} P_\textrm{L} \nu_{l})
    (\bar{\nu}_{\tau} \gamma _{\mu} P_\textrm{L} \tau)
  + \textrm{H.c.},
\end{split}
\label{stau_lagrangian}
\end{equation}
where
$G = 1.166 \times 10^{-5} \, \mathrm{GeV^{-2}}$ is the Fermi constant,
$P_{\textrm{L}}$ and $P_{\textrm{R}}$ are the chiral projection
operators,
$l \in \{\mathrm{e}, \mu\}$, and
$g_{\textrm{L}}$ and $g_{\textrm{R}}$ are the coupling constants.
These coupling constants are written in terms of the
$\mathrm{SU(2)}_{\textrm{L}}$ gauge-coupling constant $g$; when the
neutralino is bino like, for instance, they are given by
\begin{align}
  g_{\textrm{L}} & =
  \frac{g}{\sqrt{2} \cos\theta_{\textrm{W}}}
  \sin \theta_{\textrm{W}} \cos{\theta _{\tau}},
  \\
  g_{\textrm{R}} & =
  \frac{\sqrt{2}g}{\cos\theta_{\textrm{W}}}
  \sin\theta_{\textrm{W}} \sin\theta_{\tau} \mathrm{e}^{\mathrm{i} \gamma_{\tau}},
\end{align}
where $\theta_{\textrm{W}}$ is the Weinberg angle. 
The interaction Lagrangian (\ref{stau_lagrangian}) give rise to the
following decay channels (see Fig. \ref{free_diagrams}):
\begin{align}
  \tilde{\tau} & \to \tau \tilde{\chi}^{0},
  \label{eq:stau-decay-2body} \\
  \tilde{\tau} & \to \pi \nu_{\tau} \tilde{\chi}^{0},
  \label{eq:stau-decay-3body} \\
  \tilde{\tau} & \to l \nu_{l} \nu_{\tau} \tilde{\chi}^{0}.
  \label{eq:stau-decay-4body}
\end{align}
Process (\ref{eq:stau-decay-2body}) has a typical lifetime
$\mathcal{O}(10^{-20}) \mathrm{sec}$, process
(\ref{eq:stau-decay-3body}) has $(10^{-6} \, \textrm{--} \, 10^{2}) \,
\mathrm{sec}$, and (\ref{eq:stau-decay-4body}) has $(10^{2} \,
\textrm{--} \, 10^{12}) \, \mathrm{sec}$.
Since the BBN takes place $(1 \, \textrm{--} \, 100) \, \mathrm{sec}$
after the big bang, the staus will decay entirely before BBN unless
the channel (\ref{eq:stau-decay-2body}) is closed.
Our scenario therefore requires $\delta m < m_{\tau} = 1.7 \, \mathrm{GeV}$.
Note that the channel (\ref{eq:stau-decay-3body}) also closes when
$\delta m$ is less than the pion mass $m_{\pi} \simeq 140 \,
\mathrm{MeV}$.
Although the required LSP-NLSP mass difference is small compared to
the typical mass of LSP which is $\mathcal{O}(100 \, \mathrm{GeV})$,
it is preferable in attributing the dark matter (DM) to the neutralino
LSPs since it allows the LSP-NLSP coannihilation.
With this tiny $\delta m$, the neutralino naturally becomes a cold
dark matter instead of warm or hot dark matter \cite{WDM} even though it
is produced non thermally.
Hence our model is free of the constraints from the large-scale
structure formation of the Universe.

\subsection{Interactions of staus with $^7$Be and $^7$Li}
\label{interactions_of_staus}

In this section, we consider the stau-nucleus interaction processes
that are relevant to the primordial BBN.
Three processes are discussed: (1) the hadronic-current interaction,
(2) stau-catalyzed fusion, and (3)
internal conversion of stau-nucleus bound state.
We consider the lifetimes of each process because it is crucial 
to understand the impact upon the modification of BBN.

\subsubsection{Destruction of nuclei by a hadronic-current interaction
  with free staus}

Staus can interact with the nuclei through the hadronic current
and thereby alter the BBN processes.
The abundances of $\mathrm{^{7}Li}/\mathrm{^{7}Be}$ are changed by the
new decay channels:
\begin{gather}
  \stau \to \tilde{\chi}^{0} + \nu_{\tau} + \pi^{\pm},
  \label{eq:stau-decay} \\
  \pi^{+} + \mathrm{^{7}Li} \to \mathrm{^{7}Be},
  \label{eq:7Li-7Be} \\
  \pi^{-} + \mathrm{^{7}Be} \to \mathrm{^{7}Li} 
  \label{eq:7Be-7Li} \\
  \pi^{-} + \mathrm{^{7}Li} \to \mathrm{^{7}He}.
  \label{eq:7Li-7He}
\end{gather}
The process $\pi^{+} + \mathrm{^{7}He} \to \mathrm{^{7}Li}$ does not 
occur since $^7$He is very unstable, while
the pions can be either real or virtual; 
here the virtual pion should actually be regarded as a hadronic current 
propagating between the stau and the nucleus.  
The pions produced in the process (\ref{eq:stau-decay}) also change
the proton-neutron ratio and thereby change the primordial abundance
of the light elements.

\begin{figure}
\begin{center}
\includegraphics[height=50mm]{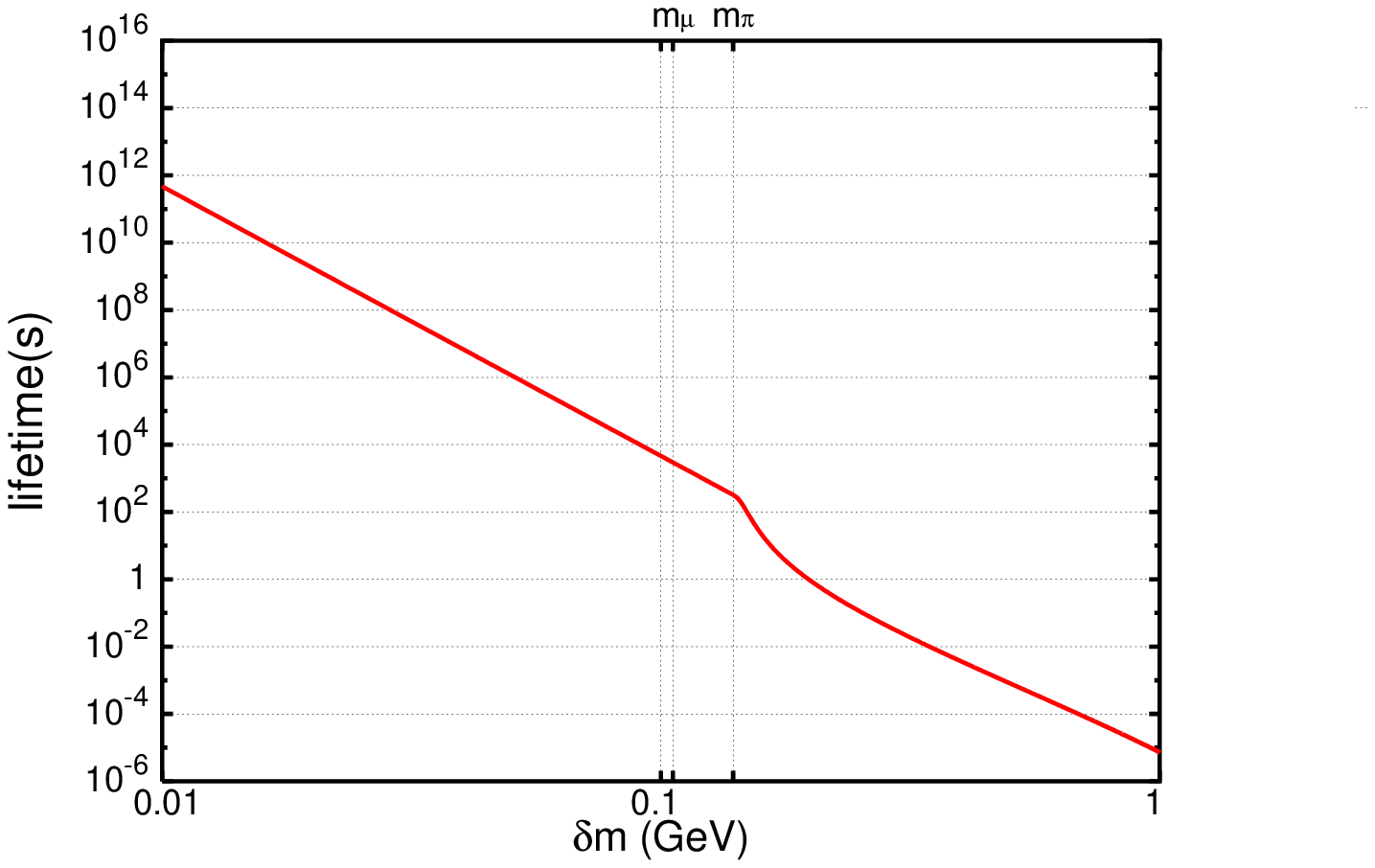}
\caption{ (color online). %
  The lifetime of free stau as the functions of $\delta m$.  Here we take
  $m_{\tilde{\chi}^{0}} = 300 \, \mathrm{GeV}$, $\theta_{\tau} =
  \pi/3$, and $\gamma_{\tau} = 0$.  
  The hadronic decay is dominant for $\delta m > m_{\pi}$ while the
  leptonic decay is exclusively allowed for $\delta m < m_{\pi}$.  }
\label{free_lifetime}
\end{center}
\end{figure}

We present the lifetime of the free stau in Fig. \ref{free_lifetime}
as functions of $\delta m$ \cite{Jittoh:2005pq}.
Here we take $m_{\tilde{\chi}^{0}} = 300 \, \mathrm{GeV}$,
$\theta_{\tau} = \pi/3$, and
$\gamma_{\tau}=0$.

\subsubsection{Stau-catalyzed fusion}

Another process to destroy the $\mathrm{^{7}Li}$/$\mathrm{^{7}Be}$ is
nuclear fusion catalyzed by staus.
A nucleus has a Coulomb barrier which normally prevents the nuclear
fusion, while the barrier is weakened when a stau is captured to a
state bound to the nucleus.
The nuclear fusion is thus promoted by forming a stau-nucleus bound
state.
The stau serves as a catalyst and is left out as the fusion proceeds
through.

This stau-catalyzed fusion process provides the following decay channels:
\begin{gather}
  \mathrm{^{7}Be} + \tilde{\tau} \to
  (\mathrm{^{7}Be} \, \tilde{\tau}) + \gamma,
  \\
  \mathrm{^{7}Li} + \tilde{\tau} \to
  (\mathrm{^{7}Li} \, \tilde{\tau}) + \gamma,
  \\
  (\mathrm{^{7}Be} \, \tilde{\tau}) + \mathrm{p} \to
  (\mathrm{^{8}B} \, \tilde{\tau}) + \gamma,
  \\
  (\mathrm{^{7}Be} \, \tilde{\tau}) + \mathrm{n} \to
  (\mathrm{^{7}Li} \, \tilde{\tau}) + \mathrm{p},
  \\
  (\mathrm{^{7}Li} \, \tilde{\tau}) + \mathrm{p}  \to
  \tilde{\tau} + 2 \, \mathrm{^{4}He} 
  \quad \textrm{or} \quad \to 
  \tilde{\tau} + 2 \, \mathrm{D} + \mathrm{^{4}He}.
\end{gather}

The lifetime of the stau-catalyzed fusion is estimated to be longer than
$1 \, \mathrm{sec}$ \cite{yanagida_catalysed}.
We follow Ref. \cite{Kawasaki:2007xb} to calculate the stau-catalyzed fusion rate.

\subsubsection{Internal conversion of nuclei in the stau-nucleus bound state}

\begin{figure}
\begin{center}
\includegraphics[height=50mm]{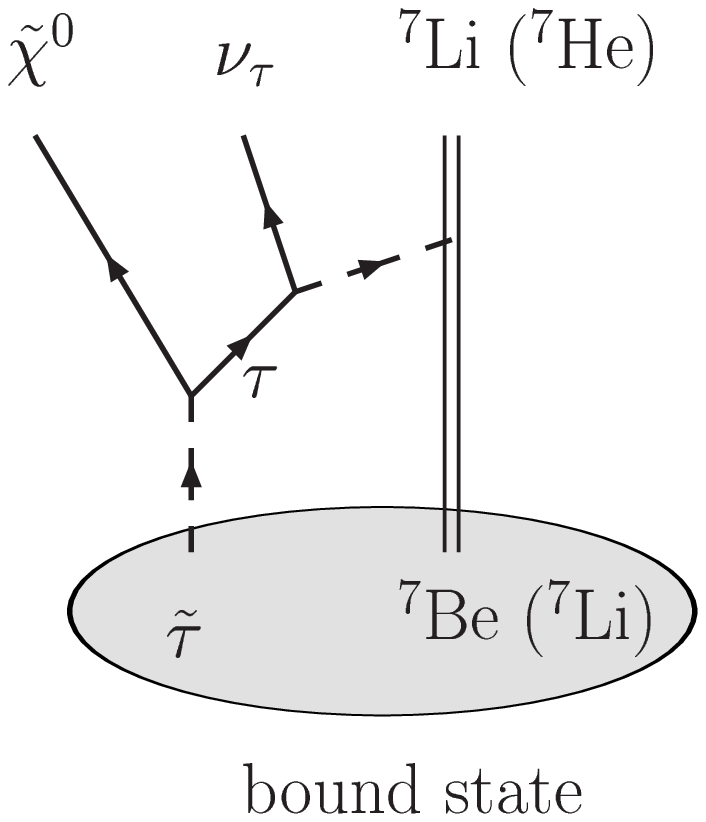}
\caption{The Feynmann diagrams of internal conversion of $^7$Be ($^7$Li).}
\label{bound_diagram}
\end{center}
\end{figure}
The interaction between a stau and a nucleus proceeds more efficiently when they
form a bound state (see Fig. \ref{bound_diagram}) \cite{Bird:2007ge}
due to two reasons: (1) the overlap of the wave functions of the
two becomes large since the stau and particle are packed in the small
space, (2) the small distance between the two allows virtual exchange of
the hadronic current even if $\delta m < m_{\pi}$.
The stau-nucleus bound state decays through the following processes:
\begin{gather}
  \tilde{\tau} + \mathrm{^{7}Be} \to
  (\tilde{\tau} \, \mathrm{^{7}Be}) \to
  \tilde{\chi}^{0} + \nu_{\tau} + \mathrm{^{7}Li},
  \label{internal_Be} \\
  \tilde{\tau} + \mathrm{^{7}Li} \to
  (\tilde{\tau} \, \mathrm{^{7}Li}) \to
  \tilde{\chi}^{0} + \nu_{\tau} + \mathrm{^{7}He},
  \label{internal_Li} \\
  \mathrm{^{7}He} \to
  \mathrm{^{6}He} + \mathrm{n},
  \label{He7_n_emission} \\
  \mathrm{^{6}He} + \textrm{background particles} \to
  \mathrm{^{3}He}, \mathrm{^{4}He}, \textrm{etc.},
  \label{He6_back_ground}
\end{gather}
where the parentheses denote the bound states.
We note that we introduce not only reaction (\ref{internal_Be}), but also reaction (\ref{internal_Li}). 
The $\mathrm{^{6}He}$ nucleus can also decay into $\mathrm{^{6}Li}$
via $\beta$ decay with the lifetime $817 \, \mathrm{msec}$.
We do not take this process into account since this process is much
slower than the scattering process (\ref{He6_back_ground}).

The lifetime of the internal conversion $\tau_{\mathrm{IC}}$ is
obtained from the Lagrangian (\ref{stau_lagrangian}) as
\begin{equation}
  \tau_{\textrm{IC}} = \frac{1}{|\psi|^2 \cdot (\sigma v)},
  \label{eq:tauIC}
\end{equation}
where $|\psi|^{2}$ is the overlap of the wave functions of the staus
and the nucleus,
\begin{equation}
\begin{split}
  (\sigma v)
  & \equiv
  \frac{1}{2E_{\tilde{\tau}}2E_{\mathrm{Be}}} 
  \int \mathrm{dLIPS} \,
  \bigl|
    \langle \neutralino \nu_{\tau} \mathrm{^{7}Li} |
    \mathcal{L}_{\textrm{int}}
    |\stau {\rm ^7Be} \rangle
  \bigr|^{2}
  \\ & \quad
  \times
  (2\pi)^{4}
  \delta^{(4)}(p_{\tilde{\tau}} + p_{\mathrm{Be}}
            - p_{\tilde{\chi^{0}}} - p_{\nu_{\tau}}-p_{\mathrm{Li}}),
  \label{eq:sigmav}
\end{split}
\end{equation}
and
\begin{equation}
  \mathrm{dLIPS}
  =
  \prod_{i} \frac{\mathrm{d}^{3} \boldsymbol{p}_i}{(2\pi)^{3} 2 E_i}.
\end{equation}
Here $i \in \{\neutralino, \nu_{\tau}, \mathrm{^7Li} \}$ for the process (\ref{internal_Be}) 
and $i \in \{\neutralino, \nu_{\tau}, \mathrm{^7He} \}$ for (\ref{internal_Li}). 
The following approximations are applied to evaluate the lifetime
further.
We estimate the overlap of the wave functions in Eq. (\ref{eq:tauIC})
by assuming that the bound state is in the S state of a hydrogen like
atom, and obtain
\begin{equation}
  |\psi|^2 = \frac{1}{\pi a_{\textrm{nucl}}^{3}},
\end{equation}
where $a_{\textrm{nucl}} = (1.2 \times A^{1/3}) \, \mathrm{fm}$ is the
radius of the nucleus.
The matrix element of the nuclear conversion appearing in
Eq. (\ref{eq:sigmav}) is evaluated by the $ft$ value of the
corresponding $\beta$ decay obtained from the experiments.
The experimental $ft$ value is available for $\mathrm{^{7}Li} \leftrightarrow
\mathrm{^{7}Be}$ but not for $\mathrm{^{7}Li} \leftrightarrow \mathrm{^{7}He}$,
however.
We assume that the two processes have the same $ft$ value.  
As long as we consider the quantum numbers of the ground 
state of Li7 and He7 we can expect a Gamow-Teller
transition can take place since they are similar with those
of He6 and Li6 and we know that they make a Gamow-Teller 
transition.  The Gamow-Teller transition is superallowed 
and has a similar $ft$ value to the Fermi transition such 
as $\mathrm{^{7}Li} \leftrightarrow \mathrm{^{7}Be}$.

Our new effects have been treated as if $^7$Li or $^7$Be in its bound 
state would have an effectively new lifetime which is caused by the 
virtual exchange of the hadronic current with a stau. Thus this new 
process is not the two-bodies scattering. So, there is no corresponding
astrophysical S factor in these processes.

\begin{figure}
\begin{center}
\includegraphics[width=80mm]{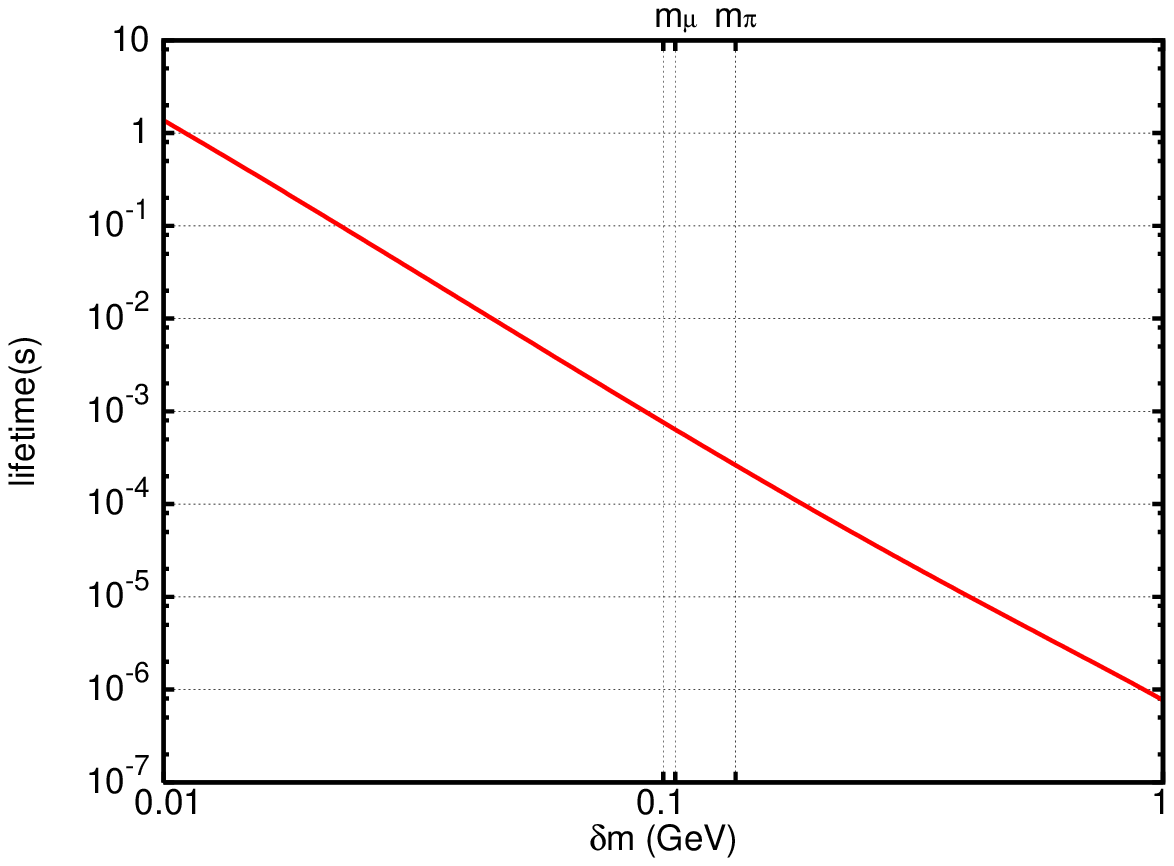}
\includegraphics[width=80mm]{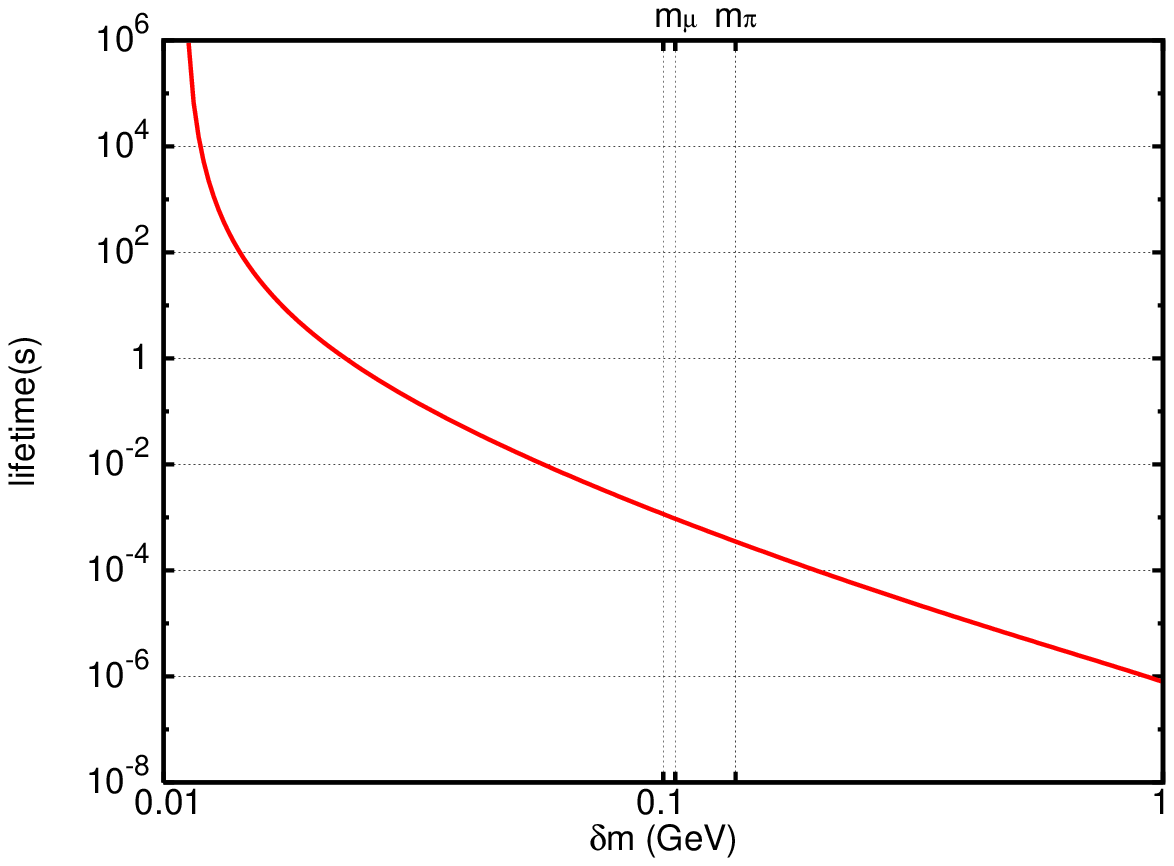}
\caption{ (color online). 
The lifetimes of internal conversion processes as the function of $\delta m$. 
Top panel: $(\stau {\rm ^7Be}) \to \neutralino + \nu_{\tau} + {\rm ^7 Li}$, 
bottom panel: $(\stau {\rm ^7Li}) \to \neutralino + \nu_{\tau} + {\rm ^7 He}$. 
We take $m_{\neutralino}$ = 300GeV, $\theta_{\tau}=\pi/3$, and $\gamma_{\tau}=0$ in both figures. 
}
\label{bound_lifetime}
\end{center}
\end{figure}

The evaluated lifetimes of reactions (\ref{internal_Be}) and
(\ref{internal_Li}) under these approximations are presented in
Fig. \ref{bound_lifetime} as functions of $\delta m$.
There we take $m_{\tilde{\chi}^{0}} = 300 \, \mathrm{GeV}$,
$\theta_{\tau} = \pi/3$, and $\gamma_{\tau} = 0$ for both reactions.
We find that the lifetime of the internal conversion process is 
in the order of $10^{-3} \, \mathrm{sec}$.
The lifetime of stau-$\mathrm{^{7}Li}$ bound state diverges around
$\delta m = m_{\mathrm{^{7}Li}} - m_{\mathrm{^{7}Be}} = 11.7 \,
\mathrm{MeV}$, below which the internal conversion is kinematically
forbidden.

As we will see later, the internal conversion processes are dominant over the other processes.

\section{Numerical calculation and interpretation of the result} 
\label{sec.numercal_calculation}

In this section, we study the effectiveness of new decay channels
on the $^7$Li problem by numerical calculation.  To do this, we
choose the abundance of stau $Y_{\stau} \equiv n_{\stau}/s|_{t=t_\mathrm{freeze~out}}$ 
at freeze out time and mass
difference $\delta m$ as free parameters, since these values are
sensitive to the abundance of $^7$Be and $^7$Li.  Here $s$ is entropy
density.  The number of $^7$Li interacted with the stau is determined by
the number density of the stau.  
The hadronic decay rate of stau is mainly determined by the mass difference.  
The decay rate is also determined by the stau mixing angle $\theta_{\tau}$, 
CP violating phase $\gamma_{\tau}$, and neutralino mass.  As we showed in
\cite{Jittoh:2005pq}, however, the effects by these parameters are
much less than mass difference.

We estimate the number density of bound states by using Saha equation, 
\begin{align}
  n_{\rm BS} = 
  \left( \frac{m_{\rm N} T}{2\pi} \right)^{-3/2} e^{E_{\rm B}/T}
  ( n_{\rm N} - n_{\rm BS} )
  ( n_{\stau^-} - n_{\rm BS} ). \label{Saha_eq}
\end{align}
Here, $n_{\rm BS}$, $n_{\stau}$, and $n_{\rm N}$ denote the number
densities of the bound state, the stau, and the nucleus, respectively;  $m_{\rm
N}$, $E_{\rm B}$, and $T$ denote the nucleus mass, binding energy of
the bound state, and the temperature of the Universe, respectively. 
The Saha equation is valid only when the expansion rate of the Universe 
is much smaller than the formation rate of the bound state.  
This condition is not quite satisfied in our case.  
We will explain that our results are qualitatively acceptable 
at the end of this section.  
For more detailed discussion, see~\cite{Kohri:2006cn,KT07}.  

\begin{figure}
\begin{center}
\includegraphics[width=85mm]{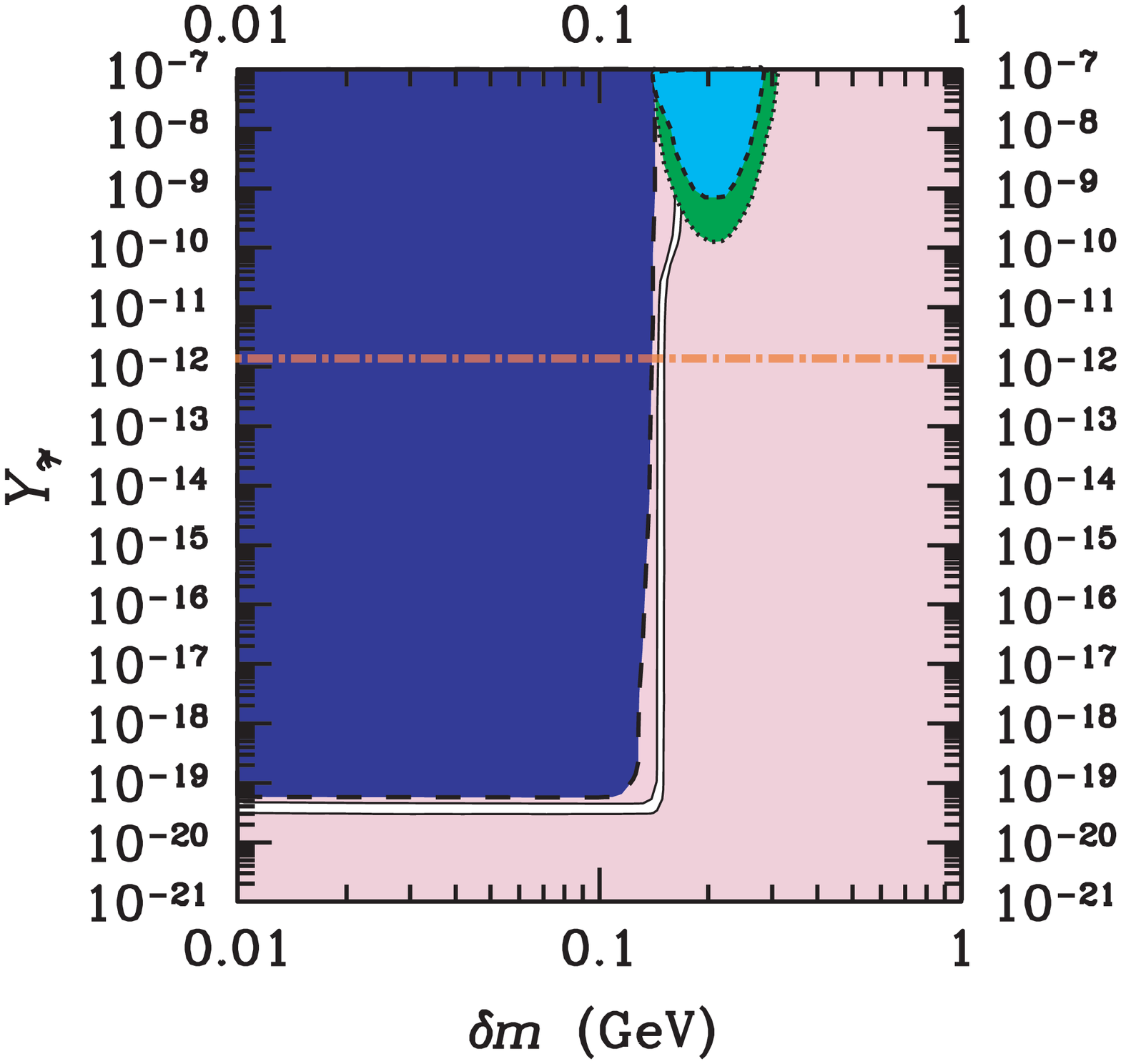}
\caption{ (color online). 
The constraints from the light-element abundance shown in the $\delta m$--$Y_{\stau}$ plane. 
The white region is the parameter space which is consistent with 
all the observational abundance including 
$^{7}$Li/H=$(1.23^{+0.32}_{-0.25})\times 10^{-10}$~\cite{Ryan:1999vr}.
The regions enclosed by dotted (green), dashed (light blue), and dash-dotted (purple) lines are excluded by the 
observations on $^4$He, D and $^{6}$Li, respectively. 
The thick-dotted line represents a yield value of stau whose daughter particle, 
neutralino, accounts for all the dark matter component. 
Here we took $\eta = 6.1\times 10^{-10}$, $m_{\neutralino}$ = 300~GeV, $\theta_{\tau}=\pi/3$
and $\gamma_{\tau}=0$. }
\label{numerical_result}
\end{center}
\end{figure}

In Fig.~\ref{numerical_result} we show the parameter region of $\delta m$
and $Y_{\stau}$ allowed by the observational light-element abundances,
where we take $\eta = 6.1\times 10^{-10}$, $m_{\neutralino}$ = 300~GeV, $\theta_{\tau}=\pi/3$
and $\gamma_{\tau}=0$.
The white region is the parameter space which is consistent with 
all the observational abundance including 
$^{7}$Li to hydrogen ratio ($^{7}$Li/H). 
The regions enclosed by dotted (green), dashed (light blue), and dash-dotted (purple) lines are excluded by the 
observations on the mass fraction of $^4$He ($Y_{\rm p}$), deuterium to  hydrogen
 ratio (D/H) and $^{6}$Li to $^{7}$Li ratio ($^{6}$Li/$^{7}$Li), respectively. 
We adopt the following observational constraints
on primordial light element abundances 
~\footnote{ Here we have used
conservative errors of $Y_{\rm p}^{\rm obs}$ according to a discussion
in Ref.~\cite{Fukugita:2006xy}.  See also the other recent
observational values of $Y_{\rm p}$ in Ref.~\cite{Peimbert:2007vm}. }
~\footnote{About the errors of $^{6}$Li/$^{7}$Li, see also the
discussion in
\cite{Pinsonneault:1998nf,Pinsonneault:2001ub,astroLi6prod,Kanzaki:2006hm}. }
\begin{eqnarray}
    \label{eq:obs}
    Y_{\rm p}^{\rm obs} &=& 0.2516 \pm 0.0040~\text{\cite{Izotov:2007ed}}, 
     \\
    \left({\rm D/H} \right)^{\rm obs} &=& (2.82
    \pm0.26)\times10^{-5}~\text{\cite{O'Meara:2006mj}}, \\ 
    {\rm Log}_{10}\left(^{7}{\rm Li}/{\rm H} \right)^{\rm
    obs} &=&-9.91 \pm 
    0.10~\text{\cite{Ryan:1999vr}} \label{Li7/H}, \\
    \left(^{6}{\rm Li}/^{7}{\rm Li}\right)^{\rm obs} &<& 0.046 \pm 0.022 +0.84
    ~\text{\cite{Asplund:2005yt}} \label{Li6/Li7}.
\end{eqnarray}
The thick-dotted line represents a yield value of stau whose daughter particle, 
neutralino, accounts for all the dark matter component. 
This line is given by the yield value of dark matter 
\begin{eqnarray}
    \label{eq:YDM}
    Y_{\rm DM} = 3.80 \times 10^{-12} 
  \left(\frac{\Omega_{\rm DM}h^{2}}{0.104}\right)
  \left(\frac{m_{\rm DM}}{10^{2} {\rm GeV}}\right)^{-1},
\end{eqnarray}
with $\Omega_{\rm DM}h^{2}=0.104 \pm 0.010$ ($68\%$
C.L.)~\cite{Spergel:2006hy}
, here $m_{\rm DM}$ is $m_{\neutralino}$.
$Y_{\tilde{\tau}}$ must be smaller than this value, in order to prevent the overclosure of the Universe.  
The cosmologically interesting region is below the line.  

We can find another white region even
if we adopt the more restrictive value of $^{7}$Li/H shown in Ref. \cite{Ryan:1999vr}.  
The upper central region is excluded by the
observational constrains on D/H and  $^{4}$He due to charged pions
emitted  from decaying staus~\cite{RStypeHad}.  In the current model,
no hadrodissociation processes of light elements occur.  

The qualitative feature of the allowed region is expalained 
from the following physical consideration.  
First, staus need to have lifetimes $\tau_{\stau}$ longer 
than the time required to form the bound state of a stau 
and a $^7$Be.  
The required time $t_{\rm form, ^{7}{\rm Be}}$ is estimated from 
the binding energy $E_{\textrm{bin}, \mathrm{^{7}Be}} = 1490 \, \mathrm{keV}$
as $t_{\rm form, ^{7}{\rm Be}} \sim 10^9 \mathrm{sec} 
\cdot (E_{\textrm{bin}, \mathrm{^{7}Be}}/\mathrm{kev})^{-2} 
\sim \mathcal{O}(10^2 \mathrm{sec})$.  
Figure \ref{free_lifetime} shows $\delta m \lesssim (100-200)\mathrm{MeV}$ 
for $\tau_{\stau} \gtrsim t_{\rm form, ^{7}{\rm Be}}$ 
and hence the allowed region appears only in this region.  
Second, the yield value of staus $Y_{\stau}$ needs to be 
large compared with that of $^7$Li which we denote by $Y_{\mathrm{^7Li}}$.  
We estimate $Y_{\mathrm{^7Li}}$ from the hydrogen to entropy 
ratio $n_\mathrm{H}/s \sim \mathcal{O}(10^{-10})$ and the constraint (\ref{Li7/H}) as 
\begin{align}
Y_{\mathrm{^7Li}} &\sim \left( \frac{n_{\mathrm{^7Li}}}{n_{\rm H}}
 \right)^{\mathrm{obs}} \cdot \frac{n_{\rm H}}{s} 
  \sim 10^{-9.63} \times 10^{-10} 
  \sim 10^{-20} .
\end{align}
We thus need $Y_{\stau} > 10^{-20}$, and again the allowed region 
appears in this region.  
Third, the excessive destruction of $^7$Li by the process (\ref{internal_Li}) 
needs to be avoided due to the constraint (\ref{Li6/Li7}).  
This condition puts a limit to the formation rate of the bound state of a stau and a $^7$Li.  
We then need either $\tau_{\stau}<t_{\rm form, ^{7}{\rm Li}} \sim 10^9 \mathrm{sec} 
\cdot (E_{\textrm{bin}, \mathrm{^{7}Li}}/\mathrm{kev})^{-2} \sim \mathcal{O}(10^3 \mathrm{sec})$ 
(here we use $E_{\textrm{bin}, \mathrm{^{7}Li}} = 952 \mathrm{keV}$ \cite{cahn:1981}), 
or $Y_{\stau}$ to be small enough.  
The former condition leads to $\delta m \gtrsim 100 \mathrm{MeV}$, 
although this region is subject to the strong restriction considered in the previous paragraph.  
The latter imposes an upper limit on $Y_{\stau}$ in the region $\delta m \lesssim 100 \mathrm{MeV}$, 
and Fig. \ref{numerical_result} suggests that this limit is in the order of $10^{-20}$.  
The tininess of $Y_{\tilde{\tau}}$ shows that the internal conversion processes (\ref{internal_Be}) and (\ref{internal_Li}) 
are dominant over other processes such as the pion exchange and the stau-catalyzed fusion.  
We can confirm this dominance by an explicit calculation of the rates of these processes.  

We used the Saha equation in our calculation although the formation rate is comparable to 
the expansion rate.  
Calculation using the Boltzmann equation will give a lower number density of the bound state 
$n_{\rm BS}$ and consequently shift the allowed region upward in $\delta m<100 \mathrm{MeV}$ 
in Fig. \ref{numerical_result}.

\section{summary} \label{sec.summary}

We have investigated a possible solution of the $^7$Li problem in a
framework of MSSM,  
in which the LSP and the NLSP are neutralino and stau, respectively, 
and have a tiny mass difference of $\delta m \lesssim (100-200) \mathrm{MeV}$. 
The staus then survive throughout the BBN era as shown in Fig. \ref{free_lifetime}, 
and provide additional decay processes as mentioned in Sec. \ref{interactions_of_staus}
to reduce the primordial $^7$Li abundance.  

Taking the three new processes into account, we numerically calculated the primordial abundance of light elements 
varying the LSP-NLSP mass difference and the abundance of stau.  
Taking $\theta_{\tau}=\pi/3,~\gamma_{\tau}=0$, we obtained the parameter region consistent to the observed $^7$Li abundance.

Though we have shown that the internal conversion is very important for the calculation of the primordial abundance, 
we need to improve our calculation for better accuracy. 
First, we need to include reaction processes such as $(\mathrm{^6Li} \tilde{\tau}) \to \mathrm{^6He} + \tilde{\chi} + \nu_{\tau}$. 
This process can change the prediction of $^6$Li and hence change the allowed region. 
Second, we should calculate the number density of the bound states not by the Saha equation (\ref{Saha_eq}) but by the Boltzmann equation. 
At the formation temperature, the capture rate is less than the expansion rate of the Universe. 
Therefore we will obtain a lower number density of the bound states 
and consequently the upward shift of the allowed region.  
Third, we should explore other values of the parameters $\theta_{\tau}$, 
$\gamma_{\tau}$, and $m_{\tilde{\chi}}$, which affect the lifetime of stau 
and also those of the bound states. 
We leave these for our future works.

\section{Acknowledgments}

J.S. thanks F. Takayama for valuable discussion.  
The work of K.K. was supported in part by  PPARC
Grant No. PP/D000394/1, EU Grant No. MRTN-CT-2006-035863, the European Union
through the Marie Curie Research and Training Network ``UniverseNet",
MRTN-CT-2006-035863. 
The work of J.S. was supported in part by the Grant-in-Aid for the Ministry of Education, Culture, Sports, Science, and Technology, Government of Japan 
(No. 17740131 and 18034001). 
The work of M.Y. was financially supported by the Sasakawa Scientific Research Grant from The Japan Science Society.  



\begin{thebibliography}{1}

\bibitem{Spergel:2006hy}
    D.~N.~Spergel et al.,
    Astrophys.\ J.\ Suppl.\ Ser.\ {\bf 170}, 377 (2007)

\bibitem{Coc:2003ce}
  A.~Coc, E.~Vangioni-Flam, P.~Descouvemont, A.~Adahchour and C.~Angulo,
  Astrophys.\ J.\  {\bf 600}, 544 (2004).

\bibitem{Ryan:1999vr}
  S.~G.~Ryan, T.~C.~Beers, K.~A.~Olive, B.~D.~Fields and J.~E.~Norris,
  Astrophys.\ J.\  Lett. {\bf 530}, L57 (2000).

\bibitem{Cyburt:2003fe}
  R.~H.~Cyburt, B.~D.~Fields and K.~A.~Olive,
  Phys.\ Lett.\ B {\bf 567}, 227 (2003).

\bibitem{Bonifacio:2002yx}
    P.~Bonifacio et al.,
    Astron.\ Astrophys.\ {\bf 390}, 91 (2002).

\bibitem{Melendez:2004ni}
  J.~Melendez and I.~Ramirez,
  Astrophys.\ J.\  {\bf 615}, L33 (2004).

\bibitem{Cyburt:2003ae}
  R.~H.~Cyburt, B.~D.~Fields and K.~A.~Olive,
  Phys.\ Rev.\ D {\bf 69}, 123519 (2004)

\bibitem{Angulo:2005mi}
  C.~Angulo {\it et al.},
  Astrophys.\ J.\  {\bf 630}, L105 (2005).

\bibitem{Korn:2006tv}
  A.~J.~Korn {\it et al.},
  Nature {\bf 442}, 657 (2006).
  [arXiv:astro-ph/0608201].



\bibitem{ichikawa-Li7}
  K.~Ichikawa and M.~Kawasaki,
  Phys.\ Rev.\ D {\bf 69}, 123506 (2004); 
  K.~Ichikawa, M.~Kawasaki and F.~Takahashi,
  Phys.\ Lett.\ B {\bf 597}, 1 (2004). 

\bibitem{Jedamzik-Li}
  K.~Jedamzik,
  Phys.\ Rev.\  D {\bf 70}, 063524 (2004); 
  K.~Jedamzik, K.~Y.~Choi, L.~Roszkowski and R.~Ruiz de Austri,
  JCAP {\bf 0607}, 007 (2006). 

\bibitem{Kohri:2005wn}
  K.~Kohri, T.~Moroi and A.~Yotsuyanagi,
  Phys.\ Rev.\  D {\bf 73}, 123511 (2006).





\bibitem{cahn:1981}
  R.~N.~Cahn and S.~L.~Glashow,
  Science {\bf 213}, 607 (1981).

\bibitem{bib:CBBNold}
  R.~N.~Boyd, K.~Takahashi, R.~J.~Perry and T.~A.~Miller,
  Science.\ Vol.\ {\bf 244}, 1450 (1989);
  A.~De Rujula, S.~L.~Glashow and U.~Sarid,
  Nucl.\ Phys.\ B {\bf 333}, 173 (1990);
  S.~Dimopoulos, D.~Eichler, R.~Esmailzadeh and G.~D.~Starkman,
  Phys.\ Rev.\ D {\bf 41}, 2388 (1990).

\bibitem{Pospelov:2006sc}
  M.~Pospelov,
  Phys.\ Rev.\ Lett.\  {\bf 98}, 231301 (2007).

\bibitem{Kohri:2006cn}
  K.~Kohri and F.~Takayama,
  Phys.\ Rev.\ D {\bf 76}, 063507 (2007).

\bibitem{bib:CBBN}
  M.~Kaplinghat and A.~Rajaraman,
  Phys.\ Rev.\  D {\bf 74}, 103004 (2006);
  R.~H.~Cyburt, J.~Ellis, B.~D.~Fields, K.~A.~Olive and V.~C.~Spanos,
  JCAP {\bf 0611}, 014 (2006);
  F.~D.~Steffen,
  AIP\ Conf.\ Proc.\ {\bf 903}, 595 (2007).

\bibitem{yanagida_catalysed}
  K.~Hamaguchi, T.~Hatsuda and T.~T.~Yanagida,
  arXiv:hep-ph/0607256.
  
\bibitem{DoubleCharge}
  D.~Fargion and M.~Khlopov,
  arXiv:hep-ph/0507087;
  D.~Fargion, M.~Khlopov and C.~A.~Stephan,
  Classical\ Quantum\ Gravity\ {\bf 23}, 7305 (2006);
  K.~M.~Belotsky, M.~Y.~Khlopov and K.~I.~Shibaev,
  Gravitation Cosmol. {\bf 12}, 93 (2006).

\bibitem{Hamaguchi:2007mp}
  K. Hamaguchi, T. Hatsuda, M. Kamimura, Y. Kino and T. T. Yanagida,
  Phys.\ Lett.\ B {\bf 650}, 268 (2007). 

\bibitem{Kawasaki:2007xb}
  M.~Kawasaki, K.~Kohri and T.~Moroi,
  Phys.\ Lett.\ B {\bf 649}, 436 (2007). 

\bibitem{Bird:2007ge}
  C.~Bird, K.~Koopmans and M.~Pospelov,
  arXiv:hep-ph/0703096.

\bibitem{Jittoh:2005pq}
  T.~Jittoh, J.~Sato, T.~Shimomura and M.~Yamanaka,
  Phys.\ Rev.\  D {\bf 73}, 055009 (2006).

\bibitem{WDM}
  W.~B.~Lin, D.~H.~Huang, X.~Zhang and R.~H.~Brandenberger,
  Phys.\ Rev.\ Lett.\  {\bf 86}, 954 (2001);
  J.~Hisano, K.~Kohri and M.~M.~Nojiri;
  Phys.\ Lett.\ B {\bf 505}, 169 (2001).

\bibitem{KT07}
  K. Kohri and F. Takayama in preparation.

\bibitem{Fukugita:2006xy}
  M.~Fukugita and M.~Kawasaki,
  Astrophys. J. {\bf 646} (2006) 691.

\bibitem{Peimbert:2007vm}
  M.~Peimbert, V.~Luridiana and A.~Peimbert,
  arXiv:astro-ph/0701580.

\bibitem{Pinsonneault:1998nf}
    M.~H.~Pinsonneault, T.~P.~Walker, G.~Steigman and V.~K.~Narayanan,
    Astrophys.\ J.\  {\bf 527}, 180 (1999).

\bibitem{Pinsonneault:2001ub}
    M.~H.~Pinsonneault, G.~Steigman, T.~P.~Walker and V.~K.~Narayanans,
    Astrophys.\ J.\  {\bf 574}, 398 (2002).
  
\bibitem{astroLi6prod}
    T.~K.~Suzuki and S.~Inoue,
    Astrophys.\ J.\ {\bf 573}, 168 (2002);
  E.~Rollinde, E.~Vangioni and K.~A.~Olive,
  Astrophys.\ J.\  {\bf 651}, 658 (2006);
  V.~Tatischeff and J.~P.~Thibaud,
  Astron. and Astrophys. in press [arXiv:astro-ph/0610756].

\bibitem{Kanzaki:2006hm}
  T.~Kanzaki, M.~Kawasaki, K.~Kohri and T.~Moroi,
  Phys.\ Rev.\  D {\bf 75}, 025011 (2007).

\bibitem{Izotov:2007ed}
  Y.~I.~Izotov, T.~X.~Thuan and G.~Stasinska,
  arXiv:astro-ph/0702072.

\bibitem{O'Meara:2006mj}
    J.~M.~O'Meara, S.~Burles, J.~X.~Prochaska, G.~E.~Prochter, 
    R.~A.~Bernstein and K.~M.~Burgess,
    Astrophys.\ J.\ {\bf 649}, L61 (2006).

\bibitem{Asplund:2005yt}
  M.~Asplund, D.~L.~Lambert, P.~E.~Nissen, F.~Primas and V.~V.~Smith,
  Astrophys.\ J.\  {\bf 644}, 229 (2006).

\bibitem{RStypeHad}
  M.~H.~Reno and D.~Seckel,
  Phys.\ Rev.\  D {\bf 37}, 3441 (1988);
  K.~Kohri,
  Phys.\ Rev.\  D {\bf 64}, 043515 (2001);
  M.~Kawasaki, K.~Kohri and T.~Moroi,
  Phys.\ Rev.\  D {\bf 71}, 083502 (2005);
  Phys.\ Lett.\  B {\bf 625}, 7 (2005);
  K.~Jedamzik,
  Phys.\ Rev.\  D {\bf 74}, 103509 (2006).

\end{thebibliography}
\end{document}